\documentclass[twocolumn,showpacs,showkeys,preprintnumbers,amsmath,amssymb]{revtex4}
\usepackage[dvips]{graphicx}% Include figure files
\usepackage{dcolumn}% Align table columns on decimal point
\usepackage{bm}% bold math

\begin{document}

\preprint{}

\title{ Study of surface potentials \\
using resonant tunneling of cold atoms in optical lattices}

\author{Vladyslav V. Ivanov}
\email{vladivanov78@gmail.com}
\affiliation{Physics Department, University of Wisconsin
Madison WI 53706, USA}

\date{\today}% It is always \today, today,

\begin{abstract}
We study a feasibility of precision measurements of surface potentials at micrometer distances using resonant tunneling
of cold atoms trapped in vertical optical lattices.
A modulation of an amplitude of
the lattice potential induces atomic tunneling among the lattice sites.
The resonant modulation frequency corresponds to a difference of potential energy between latices sites which is defined by external
force i.e. gravity. The vicinity of the surface alters the external potentials, and hence the resonant frequency.
Application of this method allows accurate study of Casimir-type potentials
and improvement of the present experimental validity of the Newtonian gravitational potential at the distance range of 5 - 30 $\mu$m.

\end{abstract}



\maketitle

\section{Introduction}

The interest in surface potentials spans from fundamental issues to the phenomenology of surface forces.
From the theoretical point of view, at length scales smaller than 100 $\mu$m QED interactions i.e. Casimir-type potentials,
such as the Casimir-Polder potential and the thermal Livshitz potential \cite{CasPol48,Lif57}, become important.
Some recent theories, going beyond the Standard Model, foresee deviations of the gravity interaction from Newton's law
\cite{AdeHecNel03,DimGer03,AntDimDva98,ArkDimDva99,Sunr04}, that occur at spatial distances and with strength, which depend
on the details of the model.
Accurate measurements of surface forces have been performed using a variety of experimental
techniques, e.g., micro-cantilevers \cite{Ger08,Chie03,BreCarRuo02,LonChaPri03}, torsion pendulum \cite{Hoy04,KapCooSwa07,Sus11}, micromechanical torsional oscillators \cite{Dec03}, atomic force microscopes \cite{CheKliMos04} and torsion balances \cite{Mas09}.
The major difficulties of such mechanical experiments are the exact
knowledge of the geometry of the setup (distance, surface roughness, etc), the precise measurement of the very
small forces involved and strong electrostatic forces between interacting bodies, with unknown strength and spatial dependence.
The latest progresses in laser cooling allow investigating surface potentials on short distances with a high spatial resolution using the micrometric size of laser cooled atomic clouds \cite{DimGer03,Car05,HarObrCor05,Obr05,WolLemCla07,Derevianko09}
In \cite{HarObrCor05,Obr05} the authors demonstrated a shift of trap frequency
of atomic Bose-Einstein condensates (BEC) induced  by temperature-dependent Livshitz potentials.
Recently new proposals were presented for measurements of an atom-surface forces using Bloch oscillations  \cite{Car05},
and sub-Hz optical atomic transitions \cite{WolLemCla07,Derevianko09}.

In this paper we study the reachable accuracy of the measurements of surface potentials with the
method based on the coherent resonant tunneling of ultra-cold atoms trapped in modulated optical lattices.
As it was demonstrated  the resonant frequency of such tunneling can be measured with a high precision, further,
it has been proposed to employ this phenomena for precision measurements, for example of the Casimir-Polder
interaction \cite{Iva08,A10,Poli2011}.
A shift of the resonant tunneling frequency measured with localized atomic samples can provide direct measurements of surface potentials.
This method can be applied for a search of possible deviations from Newtonian gravitational law at short distances.

The paper is organized as follows.
In Sec. II we give a basic theoretical description of atomic wavefunctions in tilted periodic potentials.
We briefly discuss the phenomenon of  coherent tunneling in driven lattice potentials and its possible application
for the study of surface potentials.
In Sec. III we show calculations of expected shift of the resonant frequency in vicinity of a surface
and discuss accuracy of measurements of  Casimir-type potentials. Then we discuss the expected improvements of experimental constraints
for a Yukawa-type gravitational potential.
In Sec. IV we discuss a practical implementation of the proposed method. We provide a concrete and feasible 'to-do' list for an experimental realization of the proposed method.

\section{Theoretical Background and Methods}

\subsection{Atoms in driven optical lattices}

We consider cold atoms trapped in optical lattices in presence of an external force. Optical lattices create a periodic
potential for atoms that originates from the ac Stark shift produced by the interference pattern of two counter propagating laser beams.
If the frequency of the lattice light is red detuned from the main atomic transition, the optical lattice
traps the atoms at the antinodes of the interference pattern.
Transverse confinement is provided by the Gaussian profile of the lattice beam.
The sinusoidal modulation of the power of the vertical optical lattice leads to
a tilted periodic potential in the form:

\begin{multline}
U(z,t)=mgz+\frac{U_{0}}{2}\cos[2k_{L}z](1
+\kappa\cos(\omega_{M}t)),
\label{eq:PotAm}
\end{multline}

where $mgz$ is the gravity potential, $U_{0}$ is the lattice depth,
 $\kappa$ is the modulation amplitude,
$k_{L}$ is the optical-lattice wave vector,  $\omega_{M}$ is the modulation frequency.
The recoil energy $E_{r}=\hbar^{2}k_{L}^{2}/2m$
 associated with the absorption or emission of a photon of the lattice laser is the natural energy unit for the lattice trap depth.

This potential is well studied in a static case i.e. when $\kappa=0$.
The quasi eigenstates in tilted periodic potentials are known as Wanier-Stark states.
The wavefunction of these states are localized in wells of the lattice potential.

In typical experimental conditions of vertical optical lattices these states can be considered as stationary states of the lattice potential.
Tilted periodic potential lead to the formation of a discrete ladder of energy states, the
Wannier-Stark ladder, spaced by the Bloch frequency  \cite{Wan60} in energy units.

\begin{equation}
\nu_{B}=\frac{mg\lambda}{2h},
\label{eq:BlochFreq}
\end{equation}

where $m$ is the atomic mass, $g$ is the gravity acceleration, $\lambda=2 \pi /k_{L}$ is the wavelength of the lattice light, and $h$ is the Planck constant.

The modulation of the lattice potential causes a delocalization of the atomic sates through coherent tunneling of atomic wavefunctions
among different states of the Wannier-Stark (WS) ladder. As it has been demonstrates experimentally \cite{Iva08,A10}
the coherent tunneling is a resonant process
with resonant frequencies match spacing between states of the WS ladder.
The tunneling probability amplitude as a function of the modulation frequency $\nu$ exhibits a shape
proportional to $\textbf{sinc}[(\nu-\nu_{0})/\Gamma]$ in the frequency domain based based on the generalized model of two-level system.
Here $\textbf{sinc}(x)$ is the resonance function $\textbf{sin}(x)/x$ for a
two-level transition probability, $\nu_{0}$ is the resonant frequency, $\Gamma$ is the linewidth of the transition,
and $\nu$ is the modulation frequency.
In absence of decoherence sources the linewidth  ($\Gamma$) is Fourier limited, the narrow tunneling transitions allow measurements resonant frequencies with a high precision.

\subsection{Method of studies of surface potentials}

\begin{figure}[ht]
\includegraphics[width=84mm]{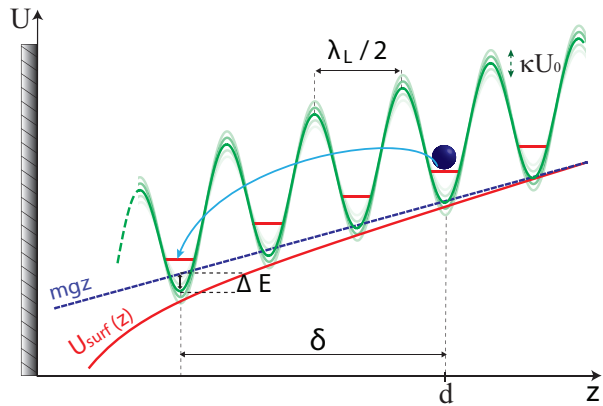}\rule{5mm}{0pt}
\caption{\label{fig:scheme} Intraband tunneling transitions in a modulated lattice potential.
The resonant frequency of these transitions is defined by energy differences of the levels in different lattice sites.
The vicinity of the surface potentials shifts the energy levels, and hence alters the resonant frequencies.}
\end{figure}

We discuss possibility to use the resonant tunneling in amplitude-modulated lattice potentials
as a probe for surface potentials, i.e. Casimir-type or non-Newtonian Yukawa-type potentials.
The basic scheme of our method is shown in  Fig.~\ref{fig:scheme}.
A sample of cold atoms is positioned close to a plane horizontal surface, and loaded into
 a vertical far-off-resonance optical lattice. The optical lattice is formed by an incoming beam and its reflection from the horizontal surface.
 Then the optical power of the lattice  beam is periodically modulated. If the modulation frequency is equal
to an energy difference of the levels of the WS ladder, coherent tunneling
 occurs between resonant lattice sites.
 For the atoms far from the surface the energy difference of the levels is set by the earth gravity.
In the vicinity of the surface the WS energy levels are shifted by the surface potentials.
This causes a shift of the resonant frequency, which depends on distances $z$ from the surface.

If site-to-site tunneling occurs on distances ($\delta$) larger than a size of the atomic sample,
the resonant tunneling causes an appearance of a separated atomic cloud. Otherwise resonant tunneling appears as a broadening of the initial atomic sample. Although a study of the surface potentials is also possible by measuring such broadening
for the sake of simplicity we focus on case in which the appearance of a separated atomic cloud is directly observed.
Here we suggest to observe a single tunneling process in contrast to \cite{Iva08,Poli2011} where multiple resonant tunneling
transitions occurs.
The surface potentials are studied by measuring the shift of the tunneling resonant frequency at different distances, and observed directly by absorption imaging.

This method can be extended to a study of possible deviations from the Newtonian gravitational law at short distances.
Expected deviations from the Newtonian gravitational potential are much weaker than the Casimir-type potential.
Thus differential measurements to exclude contribution
of the Casimir-type potential are desirable.
A study possible Yukawa-type gravitational potentials is possible using the scheme proposed in  \cite{DimGer03},
where the measurements are performed above two different materials with drastically different densities
that are covered with a thin layer of a good conductor (for example gold), forming the so-called Casimir shield.
This provides the identical Casimir potential above all surface of the sample.

\subsection{The sensitivity of the method}

%The narrow, Fourier limited tunneling resonances provide an important advantage of the described method for precision measurements.

The described method relies on accurate measurements of a number of tunneled atoms for various modulation frequencies. Fluctuations or uncertainties of these atomic numbers directly affect sensitivity. There is a few factors that might cause such fluctuations or uncertainties.
The shot noise in the atomic numbers poses fundamental limit to a signal-to-noise ratio. For example for an atomic cloud of $10^{5}$ atoms and the fraction of tunneled atoms of about $10\%$, it limits a signal-to-noise ratio to 100. Imperfections of the imaging system, such as limited numerical aperture of imaging lenses or limited quantum efficiency of CCD camera will further decrease a signal-to-noise ratio.
Instabilities of a power of lattice beams cause fluctuation of the number of tunneled atoms. Short-term instabilities for time-scale
smaller then interrogation times will be effectively averaged out during each interrogation time. A long-term instability seems more problematic.
Fluctuations of the atomic numbers in a first approximation are linear to fluctuations of the lattice beam powers. Assuming long term instability of the beam power of $10^{-3}$ and a fraction of tunneled atoms of about $10\%$ we obtain a signal-to-noise ratio of 500.

The sensitivity is directly affected by broadening of tunneling resonances. Broadening might be caused by various sources of decoherences, similar as decoherences lead to dephasing of Bloch oscillations in \cite{Gus08}. The  atomic collisions, or mean-filed interaction in case of BECs appear to be the main candidate.
As it was demonstrated \cite{Gus08} for high atomic densities a scattering length $a_{0}$ needs to be $\leq0.5a_{B}$ (where $a_{B}$ is the Bohr radius) for preservation of coherent Bloch oscillations on time scale of $\sim$10 seconds.
For thermal clouds this constraint can be far more relaxed due to lower atomic densities. A typical time between atomic collisions should be longer than an interrogation time. For example for $^{174}$Yb this implies densities in order of 10$^{10}$ cm$^{-3}$.

Fluctuations of the lattice wavelength shift resonant frequencies, which after averaging over many tunneling profiles appears as broadening.
Such fluctuations
of about 100 MHz  are reported to be main source of uncertainty in \cite{Poli2011}
 for the gravity measurement, limiting the relative uncertainty to 2$\times10^{-7}$.
Locking the
lattice laser to an atomic transition or ultra-stable cavity can decrease the uncertainty to at least a factor of 100.
Trapped atoms move in a transverse direction, thus experience different potential barriers, however this affects magnitude of the tunneling rate,
but does not affect its resonant frequency, hence does not decrease the sensitivity  \cite{Iva08}.

A modulation time is a trade-off: a longer modulation time potentially can lead to narrower resonances, however
various decoherences sources tend to limit resonance linewidth, further, atoms will be lost due to collisions with background gas.
Based on recent experiments \cite{A10,Poli2011} modulation time in order of 10 seconds appear optimum.
For example assuming the signal-to-noise ratio of 10, the modulation time of 10 seconds and an average over the 100 resonance spectrums
one can expect sensitivity in order of 0.3 mHz.
 Assuming tunneling over the 8 lattice sites this corresponds to sensitivity of measurements of
a local gravity acceleration of $\simeq$70$\times 10^{-9}$ for $^{88}$Sr and $\simeq$35$\times 10^{-9}$ for $^{174}$Yb.
Such parameters are rather modest and easily reachable with present experimental techniques.
More ambitiously one can assume the signal-to-noise ratio of 50, the interrogation time of 30 seconds and an average over the 1000 resonance spectrums
 sensitivity is expected in order of $\sim$7 $\mu$Hz. Although such parameters seem reachable with present experimental techniques they require
combination of state-of-art imaging and stabilization of lattice beams, large sample of laser-cooled atoms.

\section{Projected precision of measurements of surface potentials}

Generally Casimir-type potentials can be written as sum of two parts, where the first $U_{0}(z)$ comes from zero-point electromagnetic field fluctuations and the second $U_{th}(z)$ from the thermal fluctuations of the electromagnetic field.

\begin{equation}
U_{CP}(z)=U_{0}(z)+U_{th}(z).
\label{eq:surf}
\end{equation}

For distances much larger than
$\lambda/2\pi$, where $\lambda$ is the wavelength of strongest optical transition one can apply the so-called static approximation
to estimate the atom-surface interaction in the case atomic polizability $\alpha$ can be substituted by its static polirizability  \cite{AntPitStr05}.
The first term gives rise to the Casimir-Polder asymptotic behavior \cite{CasPol48}
\begin{equation}
U_{0}(z)=-\frac{\alpha_{0}}{4 \pi \varepsilon_{0}}\frac{3 \hbar c}{8 \pi z^{4}} \frac{\epsilon-1}{\epsilon+1} \phi(\varepsilon).
\label{eq:CP}
\end{equation}

Here $\varepsilon_{0}$ is the vacuum permittivity,
$\hbar$ is the reduced Plank constant, $c$ is the speed of light,
$\alpha_{0}$ is the static polarizability, $\varepsilon$ is static dielectric function of the substrate and $z$ is the distance
from the surface. The function $\phi(\varepsilon)$ is defined as in \cite{AntPitStr04}

The second contribution was first considered by
Lifshitz \cite{Lif57}. At large distances the thermal contribution approaches the so-called Lifshitz law

\begin{equation}
U_{th}(z)=-\frac{\alpha_{0}}{4 \pi \varepsilon}\frac{k_{B} T}{4 z^{3}} \frac{\epsilon-1}{\epsilon+1},
\label{eq:L}
\end{equation}
where $k_{B}$ is the Boltzmann constant and $T$ is the temperature of an environment.
For material with a high conductivity, such as gold or copper $\varepsilon\gg1$. This simplifies Eq.~(\ref{eq:surf}).

\begin{equation}
U_{CP}(z)=-\frac{\alpha_{0}}{4 \pi \varepsilon_{0}}\left(\frac{k_{B} T}{4 z^{3}}+\frac{3 \hbar c}{8 \pi z^{4}}\right).
\label{eq:surf2}
\end{equation}

We present non-Newtonian correction to the gravity potential in a conventional form of the Yukawa-type potential  Eq.~(\ref{eq:Yuk}).
These hypothetical corrections are usually parameterized by a Yukawa-type term which adds to the Newtonian potential \cite{AdeHecNel03,DimGer03}.
The modified potential for two masses $m_1$ and $m_2$ has the form:

\begin{equation}
V_{Yu}(r)=-G \frac{m_{1} m_{2}}{r} (1+\alpha e^{-r/\lambda_{gr}}),
\label{eq:Yuk}
\end{equation}

where $G$ is the Newton's gravitational constant, $r$ is the distance between masses,
 $\alpha$ is the relative magnitude of the Yukawa correction and $\lambda_{gr}$ its spatial range.
Assuming that the distance between the atomic cloud and the surface is much smaller that all other linear sizes (i.e. an approximation of an infinite
  plane), one obtains the Yukawa-type potential in form Eq.~(\ref{eq:YukS}).

\begin{equation}
U_{Yu}(z)=2 \pi G \rho_{0} m \alpha \lambda_{gr}^{2} e^{-z/\lambda_{gr}},
\label{eq:YukS}
\end{equation}
where $\rho_{0}$ is the bulk density of the material and $m$ is the mass of the atom.
In the approximation of an infinite plane the Newtonian gravity causes an overall shift of the potential,
but it does not introduce a potential gradient.

\subsection{Study of the Casimir-type potentials}

First we consider the effect of the Casimir-type potentials atom-surface interaction based on  Eq.~(\ref{eq:surf2}),
we discuss the feasibility of probing non-Newtonian gravitational correction later.
Two-electron atoms such as strontium or ytterbium are considered to be good candidates for precision measurements
due to their insensitivity to stray magnetic field. $^{88}$Sr is proven to be a promising candidate
also because of its negligibly small scattering length  \cite{Mar08},
which makes it immune to atomic collisions that causes decoherence.  Ytterbium atoms look advantageous for search for Yukawa-type gravitational
potential because of their large mass. Nonetheless the described method can be applied with any other laser cooled atomic species.

\begin{table}[ht]
\centerline{
\begin{tabular}{|c|c|c|c|c|c|c|c|c|c|c|c|c|}
\hline
    lattice site & 2 & 5 & 10 & 20 & 40 \\  \hline
  z ($\mu$m) & 0.40 &  1.20 & 2.53 & 5.19 & 10.5 \\  \hline
  U/(2$\pi \hbar$) (Hz) & 7.9$\times10^{3}$ &  102 & 6.5  & 0.52  & 0.050  \\  \hline
  F$_{C}/(m_{Sr}g)$ & 31.2 & 0.15 & 4 10$^{-3}$ & 160$\times10^{-6}$ & 7.2$\times10^{-6}$  \\ \hline
\end{tabular}}
\caption{ \label{fig:Table 1} Thr Casimir potentials in form Eq.~(\ref{eq:CP}) and forces for atoms of $^{88}$Sr in
different wells of the optical lattice potential.}
\end{table}

Table I shows the shift of the optical lattice potential due to the contribution of the Casimir potential
for $^{88}$Sr atoms in the optical lattice at the wavelength $\lambda_{L}=532$ nm.

\begin{figure}[ht]
\includegraphics[width=76mm]{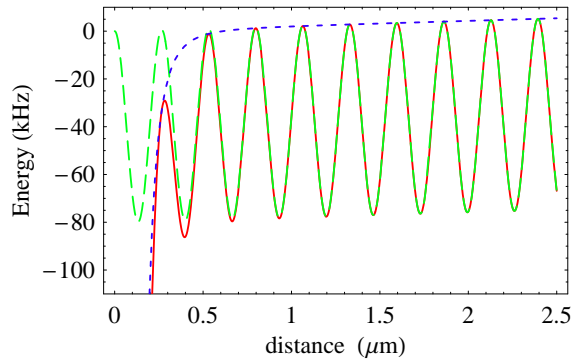}
\caption{\label{fig:Poten}
Potential energy of an atom trapped in the vertical optical lattice Eq.~(\ref{eq:PotAm}) in vicinity of a conducting surface
 at the temperature 300 K (red solid line).
The potential is calculated for $^{88}$Sr near a conducting surface in the region from 0 to 2.5 $\mu$m.
For comparison we show the lattice potential without any surface potentials (dashed green line)
and the Casimir-type potential in form of  Eq.~(\ref{eq:CP}) (blue dotted line).}
\end{figure}

\begin{figure}[ht]
\includegraphics[width=76mm]{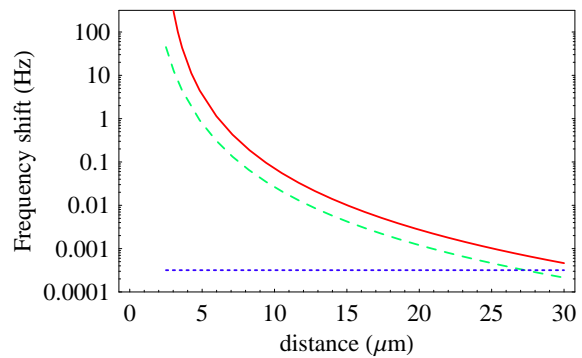}
\caption{\label{fig:CPshift}
Shift of the resonant frequency of coherent tunneling due to the Casmir-type potentials. The $^{88}$Sr atoms tunnel towards the surface.
Red solid and green dashed curves are calculated for the tunneling distances $\delta$ of 8 and  4 lattice sites correspondingly.
Horizontal blue dotted line is the level of the sensitivity of 0.3 mHz.}
\end{figure}

In Fig.~\ref{fig:Poten} we present this potential of the optical lattice in the presence of the surface potential for a lattice depth of $5E_{r}$.
The shift of the resonant frequency is determined not by overall value the Casimir-type potential but by the relative energy shift of spacing
of the WS states due to the spatially inhomogeneous surface potentials.
The frequency shift $\Delta\nu$ is proportional to the difference of the potential energy between the points $d$ and $d\pm\delta$, $\Delta\nu \sim U(d)-U(d\pm\delta)$, where $d$ is the distance from the atomic cloud to the surface, $\delta$ is the distance between the sites that are
coupled by resonant tunneling.
The expected frequency shift $\Delta\nu$ can be compared with the expected sensitivity of 0.3 mHz.
All our calculations are performed for a lattice constant of 266 nm, which corresponds to a wavelength of the lattice laser of 532 nm.
We compute the shift of the resonant frequency due to the Casimir potential  (Fig.~\ref{fig:CPshift}).
The shift of the resonance depends on the direction of the tunneling i. e. towards or away from the surface. In Fig.~\ref{fig:CPshift} and further in the paper we consider only the case of tunneling towards the surface.
The precision of the method is sensitive enough to detect change of the temperature of environment in resolution of few Kelvin degrees.

\begin{figure}[ht]
\includegraphics[width=66mm]{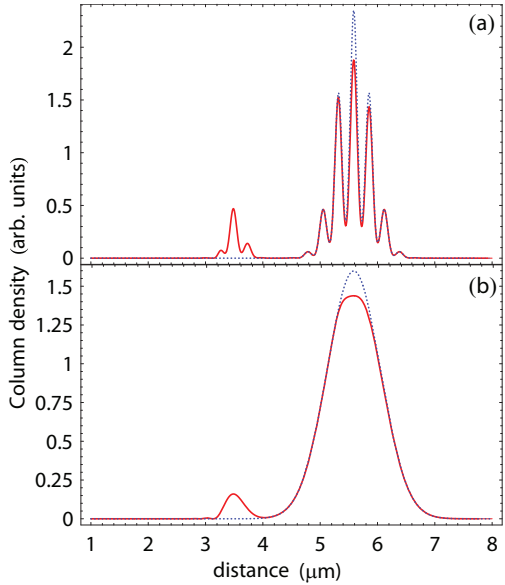}
\caption{\label{fig:ws}
Column density profile of the atomic sample trapped in the lattice potential  versus the distance from the surface.
The profiles of an atomic cloud before interrogation (blue dotted line)
and after interrogation of 10 seconds (solid red line)
are shown.
An initial distance of the sample from the surface is set to 5.5 $\mu$m,
$1/\sqrt{e}$ diameter of the sample is 1 $\mu$m.
(a) Column density profile that assumes tightly localized Wanier-Stark wavefunctions ($U_{0}=20 E_{r}$) distributed over several lattice sites
due to initial no zero diameter of atomic sample (1 $\mu$m).
(b) Column density profile assuming imaging resolution of 1 $\mu$m.
Tunneling occurs mostly from the center of the atomic sample.
Here and further we assume the resonant tunneling over 8 lattice sites.}
\end{figure}

The result of resonant tunneling through many lattice sites can be observed directly by absorption of fluorescence imaging. For illustration we demonstrate a calculated profile of the column density versus the distance from the surface in Fig.~\ref{fig:ws}. For these calculations we assume a small tunneling flux, i.e. a fraction of atoms which have tunneled twice is negligible.
Typical depth of the lattice potential is much larger that the recoil energy $E_{r}$. In this case WS wavefunctions are well localized
in each potential well. In Fig.~\ref{fig:ws}.

For small distances non-zero size of atomic samples yields a noticeable effect. Namely atoms sitting in different lattice sites
experience different potential shifts, and, hence have different resonant frequencies. Such differences has to be compared with linewidth of the transitions. For distances of 10 $\mu$m or smaller a relative shift
of potential energy for atoms siting in two neighboring are larger than the Fourier limited linewidth of 32 mHz for modulation time of 10 sec.
Basically in the frequency space strongly non-linear, curved Casimir-type potentials cause inhomogeneous broadening of tunneling profiles. Although such broadening can be computed, the sensitivity of this method relies on a small linewidth of resonances and, hence, is directly affected. Loading majority of
atoms in one lattice site would be ideal, otherwise such broadening can be a major limit for reach precision for distances smaller than 10 $\mu$m.

\subsection{Study of the Yukawa-type potentials}

\begin{figure}[ht]
\includegraphics[width=76mm]{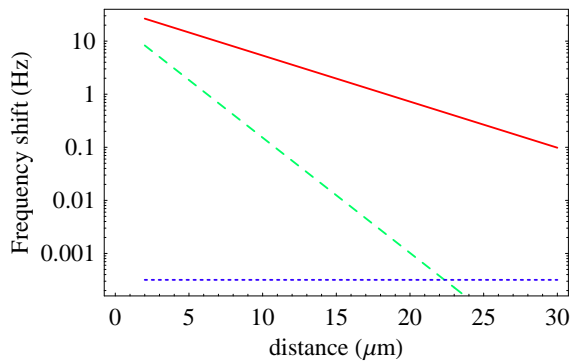}
\caption{\label{fig:YukShift}
Differential frequency shift of the resonant frequency due to the Yukawa-type gravitational potential for atoms of  $^{174}$Yb.
We choose $\lambda_{gr}$ as 1 and 5 $\mu$m, which corresponds to green dashed and red solid lines, respectively.
Dotted blue line is the expected sensitivity.
The plots are calculated for the difference of bulk densities between gold and aluminium,
parameter $\alpha$ was chosen equal to $\alpha$=10$^{9}$. }
\end{figure}

\begin{figure}[ht]
\includegraphics[width=80mm]{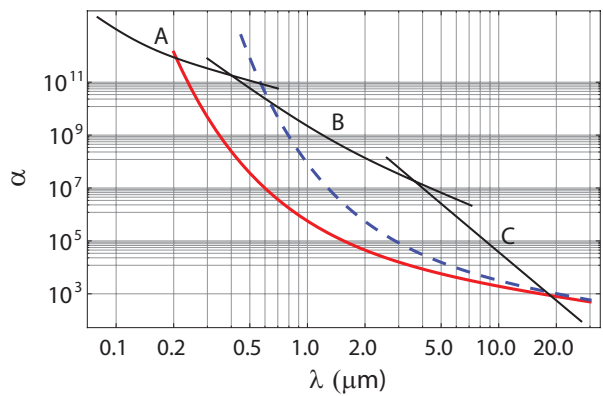}
\caption{\label{fig:ExpConstrains}
Present experimental constraints and expected experimental constraints  using our scheme in the $\alpha$-$\lambda$ diagram.
Black solid lines A, B, C show the experimental constrains achieved in  \cite{Mas09,Sus11,KapCooSwa07} correspondingly.
Non Newtonian gravity is experimentally excluded in upper-right corner of the plot.
Dashed blue and red solid lines present expected constrains achievable by present scheme for two distances of 10 $\mu$m and 5 $\mu$m
between the $^{174}$Yb atoms and the surface. }
\end{figure}

 We perform similar calculations for the frequency shift of the Yukawa-type gravitational potential.
 $^{174}$Yb  is almost twice heavier than  $^{88}$Sr, that improves projected accuracy for measurements of gravitational potentials.
 In the Fig.~\ref{fig:YukShift}  we plot the differential shift of the resonant frequency due to the Yukawa-type gravitational potential,
 based on  Eq.~(\ref{eq:YukS}),
  i.e. we plot the difference of the frequency shift above two materials with different densities for a range of the $\lambda$ parameters.
 One can evaluate reachable constraints on $\alpha$, within given sensitivity.
 To compare the expected constraints of our method on $\alpha$ with the present experimental constraints
 we use the conventional way of presentation of the experimental data, such as  $\alpha$-$\lambda$ plane  shown in Fig.~\ref{fig:ExpConstrains}.
 The left-lower corner of the plot presents  region of short $\lambda_{gr}$ and small $\alpha$,
 which needs to be investigated experimentally. Using the value of the precision 0.1 mHz we predict the experimental constraints that can be achieved
 with our scheme.
 The high spatial resolution allows a substantial improvement of the present experimental
 constraints for the short wavelengthes ($\lambda_{gr}$ $\leq$10 $\mu$m). For instance for $\lambda_{gr}=1$ $\mu$m an
 improvement on $\alpha$ of a factor of $\sim10^{3}$ is reachable for a distance of 5 $\mu$m between the atomic sample and the surface.

\section{Experimental Realization}

In order to perform the described measurements several experimental steps need to be done:\\
$\bullet$ An atomic probe, i.e. a sample of cold atoms has to be prepared. \\
$\bullet$ The atomic sample has to be accurately positioned at the needed distance from the studied surface.\\
$\bullet$ After modulation of the optical lattice one has to do readout by absorption or fluorescence imaging. \\
Here we briefly discuss the most important experimental details that might affect the accuracy of the measurements.

The presented scheme,  potentially,  can be performed with any laser cooled atomic species.
However atoms with small elastic cross section and low densities atomic samples are preferable.
Zero-spin atoms have clear advantage, since they are insensitive to magnetic fields.

Stray electric and magnetic fields
originating from contaminations of the surface reported to be significant possible error sources  \cite{HarObrCor05,Obr07,Tau2010}.
Typical local electric field strengthes caused by adsorbates are reported 100 V/cm  or less \cite{McGHarCor04,Obr07,Tau2010}
for distances of 10 $\mu$m.
A strength of electric fields caused by patch potential is expected to be the same order of magnitude or smaller
\cite{Carter2011}.
For example, for strontium a fractional shift of resonant frequency caused by electric field equals $\Delta\nu/\nu=\alpha_{Sr} E/(m_{Sr}g)$, where $\alpha_{Sr}$ is the electrostatic polarizability of Sr and $E$ is the electric field strength, and it is in order of $10^{-11}$, i.e. well below resolution of the described experiment.
Further a surface might be cleaned by heating as it was done in the Cornell group  \cite{Obr07}.

The focus of the lattice beam does not necessary coincide with a position of the atoms, that causes additional gradient of the potential
that shifts the resonant frequency. However for studied of the surface potentials in the range of 5 - 30 $\mu$m the change of the gradient
is negligible for lattice beams waist of 100 $\mu$m or larger.

Imaging poses an important problem.
The accuracy of the measurements strongly
relies on a precise determination of the distance between atoms and a surface.
 In the  recent experiment \cite{HarObrCor05} at distances of $6-10$ $\mu$m the uncertainties on $z$ are around
0.2 $\mu$m this implies a limit of about 10 \% on the measurement of $U_{CP}$.
Imaging resolution about $\sim$1 $\mu$m was achieved recently \cite{Bakr09,Hung2010}, further the short wavelengthes used for imaging of Sr and Yb (461 and 399 nm correspondingly) are advantageous.
Alternately atom-surface distances can be accurately calibrated using optical elevator as demonstrated in \cite{F08}.
The use of the retro-reflected lattice is a substantial advantage for determination of the distance from the surface.
Since the lattice phase is defined on the conducting surface by the boundary condition, the resolution of the imaging system should
be just sufficient to distinct a position of the different lattice sites. Then the exact position can be inferred.

\section{Conclusion}

In this paper we have studied feasibility of precision measurements of the surface potentials with a high spatial resolution
at distances in the range of 5 to 30 $\mu$m
using coherent resonant tunneling of atoms trapped in optical lattices.
Simple calculations show that this method can lead to more accurate measurements of Casimir-type potentials .
The search for possible deviations from the Newtonian gravitational potential is also possible. The scheme of differential measurements
should allow substantial improvements of constraints on possible non Newtonian gravity for $\lambda_{gr}$ smaller than 10 $\mu$m.

\begin{acknowledgments}
The authors thank Subhadeep Gupta and Gabriele Ferrari for critical reading of the manuscript, as well as
useful discussions about this subject. We also thank Guglielmo Tino for stimulating discussions.
\end{acknowledgments}

\end{document}